\documentclass[doublecol]{epl2}

\usepackage{graphicx} 

\usepackage{ragged2e}
\usepackage{amssymb}
\usepackage{amsmath}

\title{The mathematical journey of spatial reciprocity}
\shorttitle{The mathematical journey of spatial reciprocity}

\author{Chaoqian Wang\inst{1}\footnote{E-mail: CqWang814921147@outlook.com (corresponding author)}, Xiang Li\inst{2}, and Attila Szolnoki\inst{3}}
\shortauthor{Wang, Li, and Szolnoki}
\institute{
\inst{1}School of Mathematics and Statistics, Nanjing University of Science and Technology, Nanjing 210094, China\\
\inst{2}Yingcai Honors College, University of Electronic Science and Technology of China, Chengdu, 611731, China\\
\inst{3}Institute of Technical Physics and Materials Science, Centre for Energy Research, P.O. Box 49, H-1525 Budapest, Hungary
}

\abstract{Over the past three decades, evolutionary game theory in structured populations has witnessed explosive progress. Counterintuitive cooperative phenomena discovered in simulations have stimulated the development of rigorous mathematical theories to describe the underlying dynamics. In this paper, we review these theoretical advancements, including pair approximation and coalescent theory, which offer analytical solutions to evolutionary games on networks. We summarize the key results obtained by these theories, while also discussing their limitations, such as high computational complexity and constrained flexibility when applied to complex models. We therefore suggest that mathematical approaches provide a complementary tool for understanding cooperation in structured populations alongside simulations.}

\begin{document}

\maketitle


\section{Introduction}
Three decades ago, Nowak and May~\cite{nowak1992evolutionary} applied computer simulations to reveal unusual outcomes of spatial evolutionary games. Traditionally, defection dominates in the prisoner's dilemma game because rational individuals pursue personal gains at the expense of the collective good~\cite{nash1951non}. However, in a structured population where individuals interact only with their neighbors and copy the strategies of more successful ones, cooperation can surprisingly emerge and thrive.

Asynchronous updates via agent-based simulations~\cite{szabo1998evolutionary} have become the mainstream in subsequent research~\cite{perc2017statistical} and moved forward to the investigation of numerous coevolutionary mechanisms~\cite{perc2010coevolutionary}. These mechanisms are simple, but as a feature of complex systems, they generate unexpected behaviors whose underlying principles remain elusive. Consequently, while simulations are straightforward to implement, the profound interpretation of the emergent outcomes remains an intellectual challenge.

Among many interpretive perspectives, mathematical theories have been developed to provide analytical insights into the puzzle of cooperation in structured populations~\cite{nowak2010evolution}. These theories can identify the parameter spaces (game payoffs and network structures) under which evolution favors cooperation, though exact expressions are attainable only under special conditions. This branch is formally known as evolutionary graph theory~\cite{lieberman2005evolutionary}.

Here, we review the milestones in evolutionary graph theory over the past two decades and summarize the seminal findings. In particular, we focus on two theoretical frameworks that have attracted wide attention: the early pair approximation~\cite{ohtsuki_n06} and the more recent coalescent theory~\cite{allen_n17}. We aim to provide an overview of the achievements realized by these techniques and help readers understand the classes of problems they are equipped to solve.

\begin{figure*}[t!]
	\centering
	\includegraphics[width=.8\textwidth]{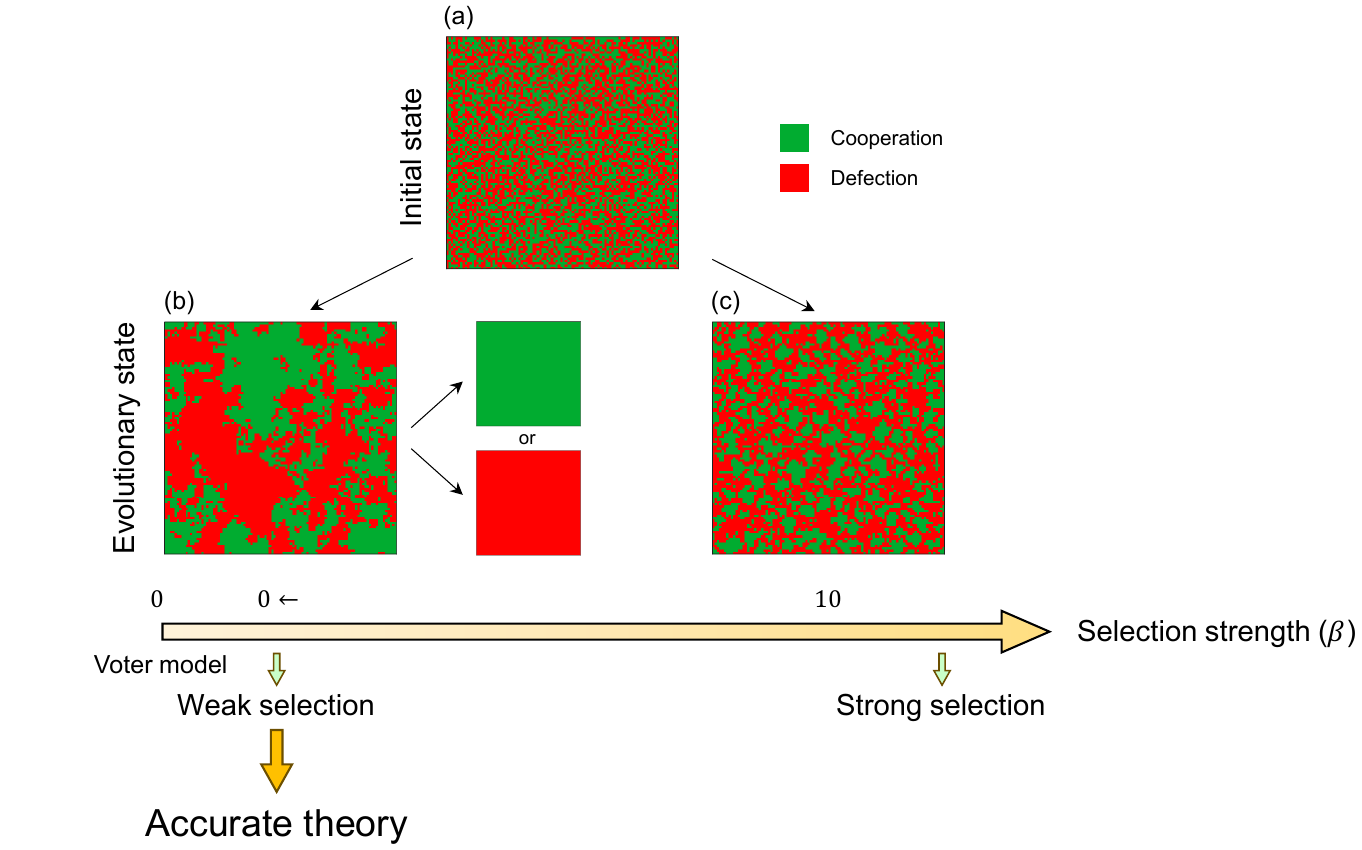}
	\caption{Typical spatial patterns of evolutionary dynamics in structured populations under different selection strengths, where the dynamics in the weak selection limit can be accurately predicted. Here, the update rule is the DB rule; the two-player game uses the $(R,S;T,P)$ payoff matrix, where $R=1$, $S=0$, $T=1.7$, and $P=0$; the population structure is a $100\times 100$ lattice with periodic boundary conditions. (a) Initial configuration with cooperators and defectors randomly distributed. (b) Under weak selection ($\beta=10^{-2}$), the evolution pattern resembles neutral drift and rapidly converges to fixation. (c) Under relatively strong selection ($\beta=10^{1}$), the evolution pattern displays a non-equilibrium stationary state observed in statistical physics.}
	\label{fig_demo}
\end{figure*}

\section{Evolutionary games on networks}
In the typical setup, individuals $i = 1, 2, \dots, N$ in a population of size $N$ are placed on a network structure. Each individual $i$ obtains an instant payoff $\pi_i$ through interactions with neighbors and updates their strategy (cooperation C or defection D) by selfish learning from neighbors. The self-interested principles can be implemented through various updating details, including pairwise comparison (PC), death-birth (DB), and imitation (IM)~\cite{ohtsuki2006replicator}. The PC rule compares the payoff with a random neighbor, which is computationally efficient in simulations. The mathematical challenge has no substantial difference across these rules, but the DB rule, where all neighbors compete for the focal position, can produce non-trivial results.

All strategy updating rules incorporate a selection strength parameter ($\beta$), the inverse of which is referred to as noise in statistical physics. Strategy updates are driven by individual fitness $F_i = \exp(\beta \pi_i)$, which linearizes to $1 + \beta \pi_i$ in the weak selection limit ($\beta \to 0$). Higher selection strength implies that individuals are more sensitive to payoffs. The patterns of strategy evolution vary across different selection strengths (Fig.~\ref{fig_demo}). For general selection strengths, the fraction of cooperative individuals, $x_\text{C}$, can fluctuate around a non-equilibrium stationary state. Analytical approaches to this scenario are constrained by exponential computational complexity~\cite{ibsen2015computational}, feasible only on special structures such as star networks~\cite{santos2008social,hadjichrysanthou2011evolutionary,mcavoy2021intriguing}. Conversely, the weak selection limit allows for perturbative methods atop the established baseline of neutral drift ($\beta = 0$), which leads to accurate solutions.

\section{Criterion for cooperation success}
\begin{figure*}[t!]
	\centering
	\includegraphics[width=\textwidth]{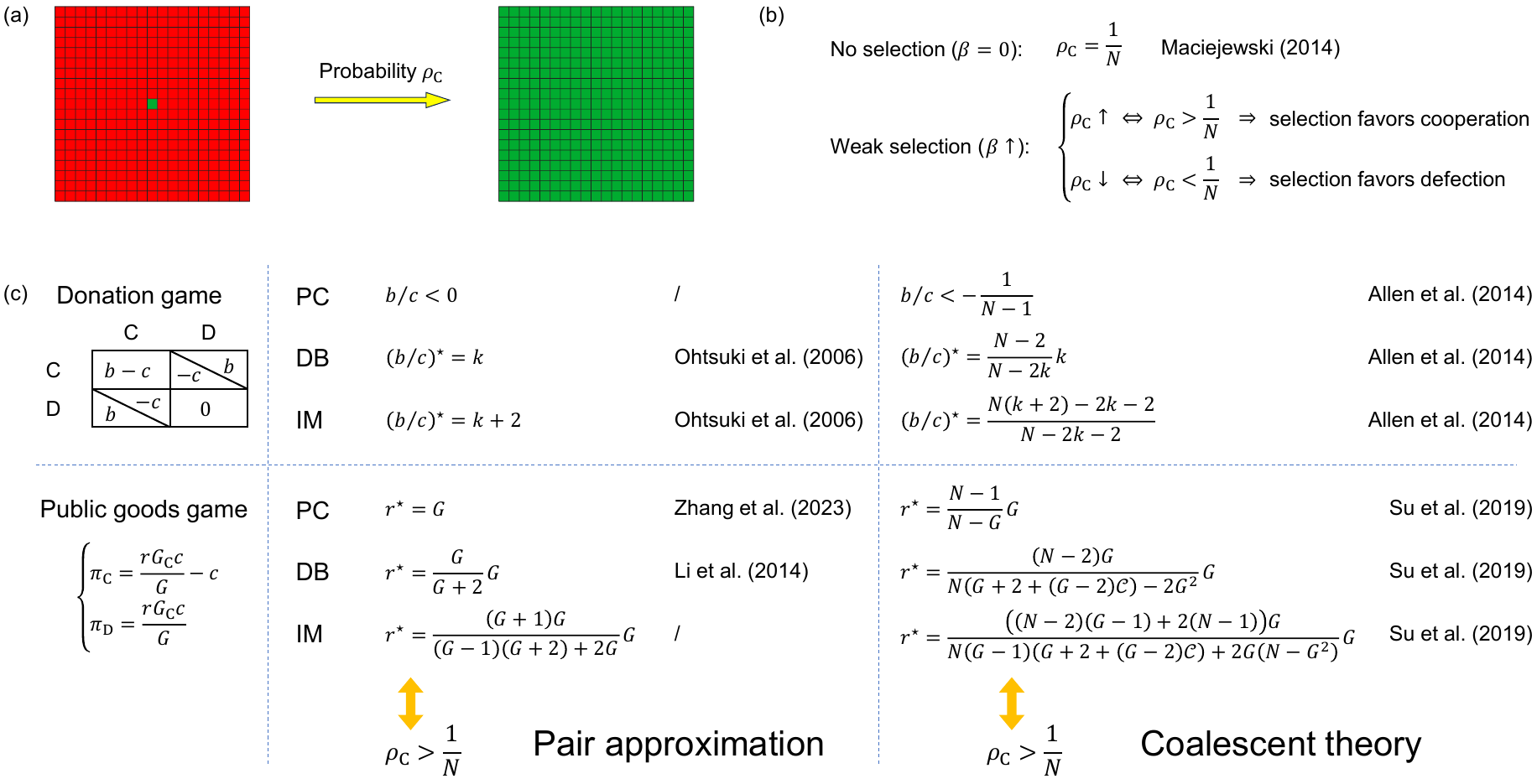}
	\caption{(a) Initialization with a randomly positioned single cooperator within an all-defector population of size $N$. The system eventually reaches fixation in all-cooperator with probability $\rho_\text{C}$ and all-defector with probability $1-\rho_\text{C}$. (b) In the absence of selection, $\rho_\text{C} = 1/N$~\cite{maciejewski2014reproductive}. When selection is introduced, if $\rho_\text{C}$ tends to increase ($\rho_\text{C} > 1/N$), the selection is said to favor cooperation. (c) Summary of theoretical conditions under which selection favors cooperation on regular graphs. We present the results for the donation game ($b/c>(b/c)^\star$) and the public goods game ($r>r^\star$) across three update rules (PC, DB, and IM) derived via pair approximation (in infinite populations) and coalescent theory (in finite populations). These results were first published by Ohtsuki {\it et al.}~\cite{ohtsuki_n06}, Li {\it et al.}~\cite{li_a_srep14}, Zhang {\it et al.}~\cite{zhang2023cooperation}, Allen {\it et al.}~\cite{allen_ems14}, and Su {\it et al.}~\cite{su_q_prsb19}. ``/'' indicates no previous publication to the best of our knowledge. See \cite{wang_cq_amc23} for the values of clustering coefficient $\mathcal{C}$ on common regular networks.}
	\label{fig_regular}
\end{figure*}

From a theoretical perspective, evolutionary dynamics in finite populations is a Markov process with only two absorbing states: $x_\text{C} = 1$ and $x_\text{C} = 0$. Under weak selection, the system can quickly reach fixation in one of them. To mathematically define cooperation success, the fixation probability of cooperators, $\rho_\text{C}$, is introduced, which represents the probability that the system eventually reaches $x_\text{C} = 1$ [Fig.~\ref{fig_regular}(a)]. This probability is numerically equivalent to the average stationary fraction of cooperation over multiple simulation runs.

In the absence of selection ($\beta=0$), the system can be understood as the Voter model~\cite{clifford1973model}, with relevant theoretical foundations dating back before the dawn of evolutionary game theory~\cite{kimura1962probability}. The general result for the fixation probability $\rho_\text{C} = \rho_\text{C}^\circ$ in this case was established by Maciejewski~\cite{maciejewski2014reproductive}: a single cooperator randomly introduced into an all-defector population leads the system to all-cooperator with probability $\rho_\text{C}^\circ = 1/N$. On heterogeneous networks, $\rho_\text{C}^\circ$ depends on the cooperator's initial position, but $\rho_\text{C}^\circ = 1/N$ still holds if the initial position is uniformly randomized.

In this way, evolutionary game models can be understood as an extended Voter model, wherein strategy evolution is influenced by payoffs. A natural criterion is thus introducing this payoff effect as a weak perturbation and observe whether $\rho_\text{C}$ tends to increase or decrease. If $\rho_\text{C}$ increases ($\rho_\text{C} > 1/N$), evolution favors cooperation; if $\rho_\text{C} < 1/N$, it favors defection [Fig.~\ref{fig_regular}(b)]. Evolutionary graph theory seeks to solve for the condition $\rho_\text{C} > 1/N$ on different networks.

Below, we review two mathematical tools for deriving $\rho_\text{C} > 1/N$: pair approximation and coalescent theory. The former is an approximate theory based on ordinary differential equations, while the latter is an exact theory based on discrete Markov processes. Their analytical solutions for the classic donation game and public goods game on regular graphs are summarized in Fig.~\ref{fig_regular}(c).

\section{Pair approximation}
In short, the pair approximation can derive the replicator equation in structured populations, which describes the evolution of $x_\text{C}$ (i.e., $\dot{x}_\text{C}=\cdots$). The required assumptions include: 1) infinite population ($N\to \infty$); 2) regular network in which each individual has the same number of neighbors $k$; 3) no triangle structure (clustering coefficient $\mathcal{C}=0$). 

In addition to $x_\text{C}$ for the traditional replicator equation, pair approximation further considers $q_{\text{C}|\text{C}}$, the probability that a cooperator adjacent to another cooperator. This establishes a minimal description for structured populations. The key insight here is that, in the weak selection limit, performing a Taylor expansion of $\dot{x}_\text{C}$ and $\dot{q}_{\text{C}|\text{C}}$ leads to $\dot{x}_\text{C}=0\times \beta^0+\cdots \times \beta^1+\mathcal{O}(\beta^2)$, $\dot{q}_{\text{C}|\text{C}}=\cdots \times\beta^0+\mathcal{O}(\beta^1)$, where $\cdots$ is nonzero. This indicates that $\dot{q}_{\text{C}|\text{C}}\gg \dot{x}_\text{C}$, meaning that $q_{\text{C}|\text{C}}$ evolves much faster than $x_\text{C}$. We can thus solve for $q_{\text{C}|\text{C}} = (k-2)x_\text{C}/(k-1) + 1/(k-1)$, such that the system state can still be represented by a single variable $x_\text{C}$.

Pair approximation reveals the celebrated ``$b/c > k$'' rule~\cite{ohtsuki_n06}, which states that in an infinite population on regular network, when applying the DB update rule in a donation game, the stability condition for the equilibrium point $x_\text{C} = 1$ is that benefit-to-cost ratio $b/c$ exceeds the number of neighbors $k$. This general framework for a two-strategy two-player game on regular graphs has applications~\cite{ohtsuki2007breaking,zhong2013evolutionary,pi2026evolutionary}. Second, Ohtsuki and Nowak~\cite{ohtsuki2006replicator} also extended the general framework to multi-strategy two-player games on regular graphs and studied its evolutionary stability~\cite{ohtsuki2008evolutionary}. Applications include the multi-strategy trust game~\cite{lim2021synergy,lim2023trust,wang2024evolution} and others~\cite{mcavoy_pcbi15}.

Third, Li {\it et al.}~\cite{li_a_pre16} proposed a general framework for two-strategy multiplayer games on regular graphs. The applications include linear and threshold public goods games~\cite{li_a_srep14,li_a_jtb15}, the $N$-person snowdrift game~\cite{sui2015evolutionary,pi2024evolutionary}, among others~\cite {han_j_c24,gao2025cooperation,zhou2018coevolution,wang_sx_cnsns19,huang_f_njp19,luo2021evolutionary,wang2022decentralized,sun2023state,zhang2023cooperation,zhu2024evolutionary}. Last, Wang {\it et al.}~\cite{wang2024evolutionary} proposed the general framework for multi-strategy multiplayer games on regular graphs and finalized this research line. Then, provided coevolutionary factors as additional strategies, any multiplayer game system can be analyzed within this framework. The applications include~\cite{dai2026evolutionary,wang2027evolutionary}.

The replicator equation on graphs can also incorporate complex mechanisms, such as game transitions~\cite{su_q_pnas19}, continuous strategies~\cite{luo2025evolutionary}, asymmetric interactions~\cite{su_q_pnas22}, novel update rules~\cite{wang2024aspiration,lin2025evolutionary}, external constraints~\cite{zhang2014evolutionary}, heterogeneous networks~\cite{li2025replicator}, and other interdisciplinary scenarios~\cite{meng_y_jrsif23,jones2024containing}.

Moreover, there is another pair approximation framework under strong selection~\cite{szabo2007evolutionary}. Applications include~\cite{szabo2002phase,chen2008interaction,xu2017extortion,khoo_srep18,hauert2024spontaneous,hauert2025phase}. However, it lacks concise analytical expressions and is not exact within the applied selection strengths.

As the name suggests, pair approximation is only an approximation based on ``pairs''~\cite{matsuda1992statistical}. Higher-order information, such as the second-order neighbors, is ignored. This explains why it does not apply to networks with triangle structures. An illustration is the public goods game on regular graphs, where, compared to the exact coalescent theory, the results of pair approximation fail to present the clustering coefficient parameter $\mathcal{C}$ [Fig.~\ref{fig_regular}(c)].

\section{Coalescent theory}
Another class of mathematical tools, which we label as ``coalescent theory'' here, has advanced rapidly in recent years. Given a specific game, network structure, and initial condition, it specifies the solution for $\rho_\text{C} > 1/N$. Such solution typically consists of game parameters on the left-hand side and network structure parameters on the right-hand side, thereby establishing the relationship between strategy evolution and network topology.

Early developments were limited to transitive regular graphs~\cite{allen_ems14,su_q_njp18,su_q_prsb19,wang_cq_njp22}, such as lattices. Far from being a limitation, we believe that these results, which apply only to regular graphs, will have lasting impacts. This is because future work focusing on new mechanisms will prefer adopting regular networks as a baseline topology to isolate irrelevant factors. Examples include the attempts to unify different update rules~\cite{debarre_nc14} through personal~\cite{wang_cq_amc23,wang_cq_pre23,wang_cq_csf23,wang_cq_pa23} and incomplete social information~\cite{wang2023imitation}.

A major breakthrough in recent years has been the generalization to any population structure~\cite{allen_n17}, under not only a single mutant~\cite{allen2019mathematical} but also arbitrary initial conditions~\cite{mcavoy2021fixation,allen2021fixation}. Methodologically, the application of this mathematical tool can be summarized into the following three steps.

Step 1): Based on the given update rule and initial condition, derive the condition for cooperation success ($\cdots>0$) under neutral drift; this condition usually includes a set of coalescent times; for example, under the PC update rule and a random single initial cooperator, the cooperation condition is derived as
\begin{equation}\label{eq_ERMC}
    \sum_{i,j\in\mathcal{N}}k_i p_{ij}\mathbb{E}^\circ_\text{RMC}[(x_i-x_j)(f_i(\mathbf{x})-f_j(\mathbf{x}))]>0.
\end{equation}
Here, $\mathcal{N}$ is the population, $k_i$ is the degree of node $i$, and $p_{ij}$ is the random walk probability from node $i$ to $j$. $x_i$ is the strategy of node $i$, and $f_i(\mathbf{x})$ is the payoff of individual $i$ under the population strategy configuration $\mathbf{x}=(x_1,x_2,\dots,x_N)$. $\mathbb{E}^\circ_\text{RMC}[\cdot]$ can be expressed in terms of coalescent times.

Step 2): Substitute the payoff function $f_i(\mathbf{x})$ into the cooperation condition in Step~1) to determine which coalescent times are required. Formulate the system of equations that governs the coalescent times ($\tau_{ij}=\mathbb{E}^\circ_\text{RMC}[x_i x_j]$ or more complicated types, depending on the integration of payoff function).

Step 3): Given the network structure, solve the system of equations in Step~2) to obtain all required coalescent times. Substitute them into the cooperation success condition and obtain the final result.

Looking forward, how can we make new innovations by this mathematical tool~\cite{allen2026understanding}? According to technical complexity, we categorize potential research avenues into three Tiers, from the simplest to the most difficult:

Tier 1: \textbf{Network application of existing cooperation conditions}. Given the update rule and game payoff, if the cooperation condition has been established in the literature, then Tier 1 is to further apply this condition to more network structures. Technically, this is the execution of Step~3). It calculates the coalescent times for the given network and plugs them into the condition. For example, Fotouhi {\it et al.}~\cite{fotouhi2018conjoining} discovered that conjoining uncooperative networks can promote cooperation. Similar extensions have also been explored on isothermal graphs~\cite{allen_ncom19} and
large modular networks~\cite{fotouhi_rsif19}.

Tier 2: \textbf{Exploration of new payoff functions}. Given the update rule, derive new cooperation condition under a new payoff function $f_i(\mathbf{x})$. Technically, this is the execution of Step~2). Since Eq.~(\ref{eq_ERMC}) is obtained under neutral drift, it is independent of games. One only needs to substitute a new payoff function $f_i(\mathbf{x})$ into Eq.~(\ref{eq_ERMC}). The payoff function $f_i(\mathbf{x})$ can be arbitrary. It can be two-player or multiplayer games, played on pairwise or higher-order networks. However, it should be noted that the complexity of $f_i(\mathbf{x})$ affects the required coalescent times. For example, in public goods games~\cite{wang2024spatial}, $f_i(\mathbf{x})$ is linear in $x_i$, so multiplying by $(x_i - x_j)$ yields at most quadratic terms, demanding two-point coalescent times $\tau_{ij}$ only, which is lucky. If the game becomes more complex, extra coalescent time types, such as $\tau_{ijl}$, may be required.

Examples of Tier 2 include social goods dilemmas~\cite{mcavoy_nhb20}, public goods games~\cite{wang2024spatial,sheng2026cooperation}, stochastic games~\cite{zhang2026effects}, nonlinear social interactions~\cite{allen2024nonlinear}, strategy evolution on higher-order network~\cite {sheng_az_ncs24,wang2026evolution} and independent interaction networks~\cite{dong2023asymmetric}.

Tier 3: \textbf{Exploration of new strategy dynamics}. Derive everything from scratch under the new strategy update dynamics (i.e., start from Step~1)). It should be noted that any new factor that changes the mathematical formalization of strategy updates under neutral drift should be categorized into this tier. Examples include strategy evolution on multilayer networks~\cite{su_q_nhb22,su2022contextualized,ma2026intraspecific,li2026strategy} and dynamic/temporal networks~\cite{su2023strategy,sheng2023evolutionary,meng2025promoting,wang2026strategy,li2026temporality}, higher-order strategy updates~\cite{wang2025emergence}, heterogeneous update rates~\cite{meng_y_nc24}, self-confidence~\cite{zeng2025evolutionary}, social learning rules~\cite{li2025collective}, and aspiration-based learning rules~\cite{su2025evolutionary}.

These three tiers form a hierarchical cascade, where a breakthrough at a higher level (e.g., Tier 3) unlocks vast space at lower levels (e.g., Tiers 2 and 1). For example, once a new strategy dynamics is established at Tier 3, one can substitute various payoff functions to operate Tier 2 innovations, obtaining the results for various games. Then, each game under each strategy dynamics can be calculated across a lot of networks to produce numerous Tier 1 papers. This research field represents a blue ocean with immense potential.

\begin{figure*}
	\centering
	\includegraphics[width=.9\textwidth]{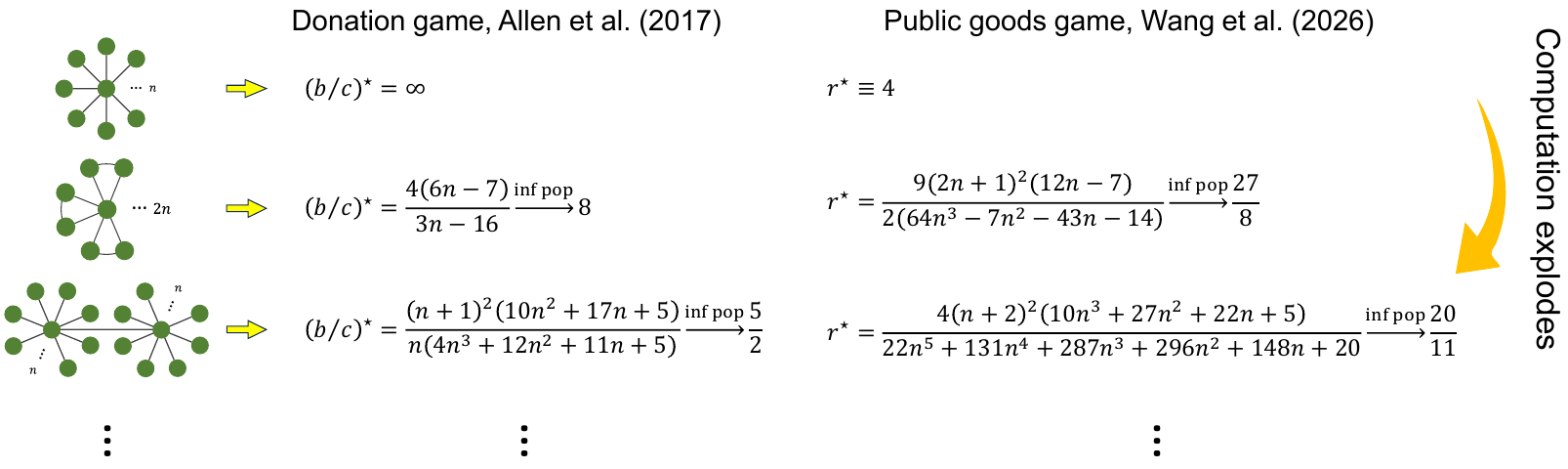}
	\caption{Exact analytical solutions on heterogeneous networks can be obtained using coalescent theory, but the computational complexity increases explosively. Here, we present results for only three simple networks under the DB update rule: a star graph with $1$ hub and $n$ leaves ($N=n+1$), a fan network with $1$ hub and $n$ two-node leaves ($N=2n+1$), and a joint star graph with $2$ hubs, each with $n$ leaves ($N=2n+2$). The exact cooperation conditions on these networks were first published by Allen {\it et al.}~\cite{allen_n17} in the donation game and by Wang {\it et al.}~\cite{wang2024spatial} in the public goods game.}
	\label{fig_any}
\end{figure*}

\section{Discussion}
In summary, the past two decades have seen remarkable progress in theoretical developments of evolutionary games in structured populations. Following early detours~\cite{lieberman2005evolutionary,taylor2007evolution}, the structure coefficient theorems~\cite {tarnita_jtb09,tarnita2011multiple,mcavoy2022evaluating,wu2013dynamic,mcavoy2016structure}, and many other techniques~\cite{su_q_njp16,su_q_njp17,su2018evolution,su2019edge,svoboda2024density,wang2024deterministic,wang2025evolutionary,xia2025behavioral}, the coalescent theory finally laid the foundation for exact analysis. Spatial reciprocity, once understood through physical intuition, can now be analyzed mathematically under specific assumptions. These assumptions include weak selection, fixation, and the simplicity of models.

This perspective summarizes the closed-form solutions for classic evolutionary game models that may interest our readers (Fig.~\ref{fig_regular}). Although the derivations of these results are purely mathematical, we look forward to interpreting the intuitive physical insights of each step in the future. At the same time, theoretical researchers are attempting to extend analytical solutions to strong selection~\cite{ohtsuki2006evolutionary,altrock2017evolutionary,wu2023evolutionary}, strategy frequencies~\cite{mcavoy2022evaluating}, and co-evolutionary mechanisms~\cite{shao2024value,gao2024evolutionary}.

One vision for theoretical developments is to bypass agent-based simulations, because the latter demand high computational resources. In the weak selection limit, the computational complexity of analytical solutions (related to an $N^2\times N^2$ matrix in solving coalescent times) scales polynomially with population size $N$. Furthermore, since the adjacency matrices of most networks are sparse, approximate algorithm can be employed~\cite{koutis2014approaching}. Even so, the computation remains substantial (exemplified in Fig.~\ref{fig_any}). To the best of our knowledge, when solving for populations of $N\sim 10^3$, the memory capacity of a personal laptop is insufficient, and high-performance servers are required. The maximum population size reported in the literature remains on $N \sim 5\times 10^3$~\cite{allen_n17}. In contrast, personal laptops can easily support agent-based simulations of large populations, such as a $400\times 400=1.6\times 10^5$ lattice.

In the age of advancing computer technology, simulations will become even increasingly efficient in the future. Here, the merit of analytical theory lies in its ability to inform researchers about which classes of models can already be theoretically predicted, and which mechanisms, even those that appear very simple, can only be resolved through simulations.

One of the most fascinating features of spatial evolutionary games is the intuitiveness and counterintuitiveness in physical imagery on large scales. The interpretation of various evolutionary outcomes is, in many ways, an art. Current developments in mathematical theory focus on statistics of local configurations but overlook the large-scale temporal and spatial effects~\cite{szolnoki_prx13,roca2009evolutionary,szolnoki_njp16}. We believe that an interesting future direction is to weave mathematical theories into the art of physical interpretation.

Every complex system has its uniqueness. Countless efforts have been made to explain the rise of cooperation through reductionist approaches~\cite{fu2009evolutionary,zhang2011ambivalent,jiang2011effects,wang2023high}, yet no universal theory can close the book on this exploration. All theories are tools with certain limitations that complement our understanding of cooperation. We therefore hope this perspective encourages fellow travelers to continue the mathematical journey of spatial reciprocity.

\begin{acknowledgments}
CW was supported by the Fundamental Research Funds for the Central Universities (30926010516) and the Stability Support Program for Basic Research (WZJC202601001). AS was supported by the National Research, Development and Innovation Office (NKFIH) under Grant No.~K142948.
\end{acknowledgments}


\begin{thebibliography}{100}
\expandafter\ifx\csname url\endcsname\relax\def\url#1{\texttt{#1}}\fi

\bibitem{nowak1992evolutionary}
\Name{Nowak M.~A. \and May R.~M.} \REVIEW{Nature}{359}{1992}{826}.

\bibitem{nash1951non}
\Name{Nash J.~F.} \REVIEW{Ann. Math.}{54}{1951}{286–295}.

\bibitem{szabo1998evolutionary}
\Name{Szab{\'o} G. \and T{\H{o}}ke C.} \REVIEW{Phys. Rev. E}{58}{1998}{69}.

\bibitem{perc2017statistical}
\Name{Perc M., Jordan J.~J., Rand D.~G., Wang Z., Boccaletti S. \and Szolnoki A.} \REVIEW{Phy. Rep.}{687}{2017}{1}.

\bibitem{perc2010coevolutionary}
\Name{Perc M. \and Szolnoki A.} \REVIEW{BioSystems}{99}{2010}{109}.

\bibitem{nowak2010evolution}
\Name{Nowak M.~A., Tarnita C.~E. \and Wilson E.~O.} \REVIEW{Nature}{466}{2010}{1057}.

\bibitem{lieberman2005evolutionary}
\Name{Lieberman E., Hauert C. \and Nowak M.~A.} \REVIEW{Nature}{433}{2005}{312}.

\bibitem{ohtsuki_n06}
\Name{Ohtsuki H., Hauert C., Lieberman E. \and Nowak M.~A.} \REVIEW{Nature}{441}{2006}{502}.

\bibitem{allen_n17}
\Name{Allen B., Lippner G., Chen Y.-T., Fotouhi B., Momeni N., Yau S.-T. \and Nowak M.~A.} \REVIEW{Nature}{544}{2017}{227}.

\bibitem{ohtsuki2006replicator}
\Name{Ohtsuki H. \and Nowak M.~A.} \REVIEW{J. Theor. Biol.}{243}{2006}{86}.

\bibitem{ibsen2015computational}
\Name{Ibsen-Jensen R., Chatterjee K. \and Nowak M.~A.} \REVIEW{Proc. Natl. Acad. Sci. U.S.A.}{112}{2015}{15636}.

\bibitem{santos2008social}
\Name{Santos F.~C., Santos M.~D. \and Pacheco J.~M.} \REVIEW{Nature}{454}{2008}{213}.

\bibitem{hadjichrysanthou2011evolutionary}
\Name{Hadjichrysanthou C., Broom M. \and Rycht{\'a}{\v{r}} J.} \REVIEW{Dyn. Games Appl.}{1}{2011}{386}.

\bibitem{mcavoy2021intriguing}
\Name{McAvoy A., Rao A. \and Hauert C.} \REVIEW{PLoS Comput. Biol.}{17}{2021}{e1009611}.

\bibitem{clifford1973model}
\Name{Clifford P. \and Sudbury A.} \REVIEW{Biometrika}{60}{1973}{581}.

\bibitem{kimura1962probability}
\Name{Kimura M.} \REVIEW{Genetics}{47}{1962}{713}.

\bibitem{maciejewski2014reproductive}
\Name{Maciejewski W.} \REVIEW{J. Theor. Biol.}{340}{2014}{285}.

\bibitem{li_a_srep14}
\Name{Li A., Wu B. \and Wang L.} \REVIEW{Sci. Rep.}{4}{2014}{5536}.

\bibitem{zhang2023cooperation}
\Name{Zhang W. \and Brandes U.} \REVIEW{Appl. Math. Comput.}{438}{2023}{127604}.

\bibitem{allen_ems14}
\Name{Allen B. \and Nowak M.~A.} \REVIEW{EMS Surv. Math. Sci.}{1}{2014}{113}.

\bibitem{su_q_prsb19}
\Name{Su Q., Li A., Wang L. \and Stanley H.~E.} \REVIEW{Proc. R. Soc. B}{286}{2019}{20190041}.

\bibitem{wang_cq_amc23}
\Name{Wang C. \and Szolnoki A.} \REVIEW{Appl. Math. Comput.}{449}{2023}{127941}.

\bibitem{ohtsuki2007breaking}
\Name{Ohtsuki H., Nowak M.~A. \and Pacheco J.~M.} \REVIEW{Phys. Rev. Lett.}{98}{2007}{108106}.

\bibitem{zhong2013evolutionary}
\Name{Zhong W., Liu J. \and Zhang L.} \REVIEW{Biosystems}{111}{2013}{102}.

\bibitem{pi2026evolutionary}
\Name{Pi B., Feng M., Deng L.-J., Chen X. \and Szolnoki A.} \REVIEW{IEEE Trans. Cybern.}{56}{2026}{4559}.

\bibitem{ohtsuki2008evolutionary}
\Name{Ohtsuki H. \and Nowak M.~A.} \REVIEW{J. Theor. Biol.}{251}{2008}{698}.

\bibitem{lim2021synergy}
\Name{Lim I.~S. \and Capraro V.} \REVIEW{IEEE Trans. Circuits Syst. II: Express Br.}{69}{2021}{2777}.

\bibitem{lim2023trust}
\Name{Lim I.~S. \and Masuda N.} \REVIEW{IEEE Trans. Evol. Comput.}{28}{2023}{117}.

\bibitem{wang2024evolution}
\Name{Wang C.} \REVIEW{Appl. Math. Comput.}{471}{2024}{128595}.

\bibitem{mcavoy_pcbi15}
\Name{McAvoy A. \and Hauert C.} \REVIEW{PLoS Comput. Biol.}{11}{2015}{e1004349}.

\bibitem{li_a_pre16}
\Name{Li A., Broom M., Du J. \and Wang L.} \REVIEW{Phys. Rev. E}{93}{2024}{0022407}.

\bibitem{li_a_jtb15}
\Name{Li A. \and Wang L.} \REVIEW{J. Theor. Biol.}{377}{2015}{57}.

\bibitem{sui2015evolutionary}
\Name{Sui X., Cong R., Li K. \and Wang L.} \REVIEW{Phys. Lett. A}{379}{2015}{2922}.

\bibitem{pi2024evolutionary}
\Name{Pi J., Wang C., Zhou D., Tang W. \and Yang G.} \REVIEW{Chaos Solitons Fractals}{180}{2024}{114593}.

\bibitem{han_j_c24}
\Name{Han J., Chen X. \and Szolnoki A.} \REVIEW{Chaos}{34}{2024}{033124}.

\bibitem{gao2025cooperation}
\Name{Gao S., Suo J. \and Li N.} \REVIEW{Appl. Math. Comput.}{489}{2025}{129156}.

\bibitem{zhou2018coevolution}
\Name{Zhou L., Li A. \and Wang L.} \REVIEW{J. Theor. Biol.}{440}{2018}{32}.

\bibitem{wang_sx_cnsns19}
\Name{Wang S., Chen X. \and Szolnoki A.} \REVIEW{Commun. Nonlinear Sci. Numer. Simulat.}{79}{2019}{104914}.

\bibitem{huang_f_njp19}
\Name{Huang F., Chen X. \and Wang L.} \REVIEW{New J. Phys.}{21}{2019}{063013}.

\bibitem{luo2021evolutionary}
\Name{Luo Q., Liu L. \and Chen X.} \REVIEW{Physica D}{424}{2021}{132943}.

\bibitem{wang2022decentralized}
\Name{Wang S., Chen X., Xiao Z. \and Szolnoki A.} \REVIEW{Appl. Math. Comput.}{431}{2022}{127308}.

\bibitem{sun2023state}
\Name{Sun Z., Chen X. \and Szolnoki A.} \REVIEW{IEEE Trans. Netw. Sci. Eng.}{10}{2023}{3975}.

\bibitem{zhu2024evolutionary}
\Name{Zhu W., Wang X., Wang C., Liu L., Hu J., Zheng Z., Tang S., Zheng H. \and Dong J.} \REVIEW{Commun. Phys.}{7}{2024}{377}.

\bibitem{wang2024evolutionary}
\Name{Wang C., Perc M. \and Szolnoki A.} \REVIEW{Nat. Commun.}{15}{2024}{5349}.

\bibitem{dai2026evolutionary}
\Name{Dai X., Liu H., Gao S., Bai H., Jia D. \and Shen C.} \REVIEW{Chaos Solitons Fractals}{208}{2026}{118250}.

\bibitem{wang2027evolutionary}
\Name{Wang Q. \and Chen X.} \REVIEW{Appl. Math. Comput.}{532}{2027}{130239}.

\bibitem{su_q_pnas19}
\Name{Su Q., McAvoy A., Wang L. \and Nowak M.~A.} \REVIEW{Proc. Natl. Acad. Sci. U.S.A.}{116}{2019}{25398}.

\bibitem{luo2025evolutionary}
\Name{Luo J., Lin D., Chen X. \and Szolnoki A.} \REVIEW{Chaos}{35}{2025}{043115}.

\bibitem{su_q_pnas22}
\Name{Su Q., Allen B. \and Plotkin J.~B.} \REVIEW{Proc. Natl. Acad. Sci. U.S.A.}{119}{2022}{e2113468118}.

\bibitem{wang2024aspiration}
\Name{Wang Q., Chen X., He N. \and Szolnoki A.} \REVIEW{IEEE Trans. Neural Netw. Learn. Syst.}{36}{2025}{8387}.

\bibitem{lin2025evolutionary}
\Name{Lin L., Li C. \and Chen X.} \REVIEW{Chaos}{35}{2025}{023113}.

\bibitem{zhang2014evolutionary}
\Name{Zhang H., Gao M., Wang W. \and Liu Z.} \REVIEW{J. Theor. Biol.}{358}{2014}{122}.

\bibitem{li2025replicator}
\Name{Li J., Wang X., Li C. \and Zhang B.} \REVIEW{J. Math. Biol.}{90}{2025}{16}.

\bibitem{meng_y_jrsif23}
\Name{Meng Y., Broom M. \and Li A.} \REVIEW{J. R. Soc. Interface}{20}{2023}{20230295}.

\bibitem{jones2024containing}
\Name{Jones M.~I., Pauls S.~D. \and Fu F.} \REVIEW{PNAS Nexus}{3}{2024}{pgae090}.

\bibitem{szabo2007evolutionary}
\Name{Szab{\'o} G. \and F{\'a}th G.} \REVIEW{Phys. Rep.}{446}{2007}{97}.

\bibitem{szabo2002phase}
\Name{Szab{\'o} G. \and Hauert C.} \REVIEW{Phys. Rev. Lett.}{89}{2002}{118101}.

\bibitem{chen2008interaction}
\Name{Chen X., Fu F. \and Wang L.} \REVIEW{Phys. Rev. E}{78}{2008}{051120}.

\bibitem{xu2017extortion}
\Name{Xu X., Rong Z., Wu Z.-X., Zhou T. \and Tse C.~K.} \REVIEW{Phys. Rev. E}{95}{2017}{052302}.

\bibitem{khoo_srep18}
\Name{Khoo T., Fu F. \and Pauls S.} \REVIEW{Sci. Rep.}{8}{2018}{6922}.

\bibitem{hauert2024spontaneous}
\Name{Hauert C. \and Szab{\'o} G.} \REVIEW{PNAS Nexus}{3}{2024}{pgae326}.

\bibitem{hauert2025phase}
\Name{Hauert C. \and Szab{\'o} G.} \REVIEW{Phys. Rev. E}{111}{2025}{054308}.

\bibitem{matsuda1992statistical}
\Name{Matsuda H., Ogita N., Sasaki A. \and Sat{\=o} K.} \REVIEW{Prog. Theor. Phys.}{88}{1992}{1035}.

\bibitem{su_q_njp18}
\Name{Su Q., Wang L. \and Stanley H.~E.} \REVIEW{New J. Phys.}{20}{2018}{103030}.

\bibitem{wang_cq_njp22}
\Name{Wang C. \and Szolnoki A.} \REVIEW{New J. Phys.}{24}{2022}{123030}.

\bibitem{debarre_nc14}
\Name{D{\'e}barre F., Hauert C. \and Doebeli M.} \REVIEW{Nat. Commun.}{5}{2014}{3409}.

\bibitem{wang_cq_pre23}
\Name{Wang C. \and Szolnoki A.} \REVIEW{Phys. Rev. E}{107}{2023}{024303}.

\bibitem{wang_cq_csf23}
\Name{Wang C., Zhu W. \and Szolnoki A.} \REVIEW{Chaos, Solit. and Fract.}{173}{2023}{113667}.

\bibitem{wang_cq_pa23}
\Name{Wang C., Zhu W. \and Szolnoki A.} \REVIEW{Physica A}{625}{2023}{129033}.

\bibitem{wang2023imitation}
\Name{Wang X., Zhou L., McAvoy A. \and Li A.} \REVIEW{Nat. Commun.}{14}{2023}{7453}.

\bibitem{allen2019mathematical}
\Name{Allen B. \and McAvoy A.} \REVIEW{J. Math. Biol.}{78}{2019}{1147}.

\bibitem{mcavoy2021fixation}
\Name{McAvoy A. \and Allen B.} \REVIEW{J. Math. Biol.}{82}{2021}{14}.

\bibitem{allen2021fixation}
\Name{Allen B., Sample C., Steinhagen P., Shapiro J., King M., Hedspeth T. \and Goncalves M.} \REVIEW{PLoS Comput. Biol.}{17}{2021}{e1008695}.

\bibitem{allen2026understanding}
\Name{Allen B.} \REVIEW{Nat. Comput. Sci.}{6}{2026}{112}.

\bibitem{fotouhi2018conjoining}
\Name{Fotouhi B., Momeni N., Allen B. \and Nowak M.~A.} \REVIEW{Nat. Hum. Behav.}{2}{2018}{492}.

\bibitem{allen_ncom19}
\Name{Allen B., Lippner G. \and Nowak M.~A.} \REVIEW{Nat. Commun.}{10}{2019}{5107}.

\bibitem{fotouhi_rsif19}
\Name{Fotouhi B., Momeni N., Allen B. \and Nowak M.~A.} \REVIEW{J. R. Soc. Interface}{16}{2019}{20180677}.

\bibitem{wang2024spatial}
\Name{Wang C. \and Su Q.} \REVIEW{Sci. Adv.}{12}{2026}{eaeb1263}.

\bibitem{mcavoy_nhb20}
\Name{McAvoy A., Allen B. \and Nowak M.~A.} \REVIEW{Nat. Hum. Behav.}{4}{2020}{819}.

\bibitem{sheng2026cooperation}
\Name{Sheng A., Su Q., McAvoy A., Wang L. \and Plotkin J.~B.} \REVIEW{Nature}{655}{2026}{1217}.

\bibitem{zhang2026effects}
\Name{Zhang Y., Feng M., Li Q., Perc M. \and Szolnoki A.} \REVIEW{Commun. Phys.}{9}{2026}{101}.

\bibitem{allen2024nonlinear}
\Name{Allen B., Khwaja A.-R., Donahue J.~L., Kelly T.~J., Hyacinthe S.~R., Proulx J., Lattanzio C., Dementieva Y.~A. \and Sample C.} \REVIEW{PNAS Nexus}{3}{2024}{pgae131}.

\bibitem{sheng_az_ncs24}
\Name{Sheng A., Su Q., Wang L. \and Plotkin J.~B.} \REVIEW{Nat. Comput. Sci.}{4}{2024}{274}.

\bibitem{wang2026evolution}
\Name{Wang Y. \and Gao S.} \REVIEW{Chaos Solitons Fractals}{204}{2026}{117776}.

\bibitem{dong2023asymmetric}
\Name{Dong G., Li S., Du R. \and Su Q.} \REVIEW{Chaos Solitons Fractals}{173}{2023}{113758}.

\bibitem{su_q_nhb22}
\Name{Su Q., McAvoy A., Mori Y. \and Plotkin J.~B.} \REVIEW{Nat. Hum. Behav.}{6}{2022}{338}.

\bibitem{su2022contextualized}
\Name{Su Q., McAvoy A. \and Plotkin J.~B.} \REVIEW{Sci. Adv.}{8}{2022}{eabm6066}.

\bibitem{ma2026intraspecific}
\Name{Ma Y., Zhang H. \and Wang C.} \REVIEW{J. Math. Biol.}{93}{2026}{7}.

\bibitem{li2026strategy}
\Name{Li M., Shi Z., Xie J., Su Q. \and Duan G.} \REVIEW{Cell Rep. Phys. Sci.}{7}{2026}{103440}.

\bibitem{su2023strategy}
\Name{Su Q., McAvoy A. \and Plotkin J.~B.} \REVIEW{Nat. Comput. Sci.}{3}{2023}{763}.

\bibitem{sheng2023evolutionary}
\Name{Sheng A., Li A. \and Wang L.} \REVIEW{PLoS Comput. Biol.}{19}{2023}{e1011333}.

\bibitem{meng2025promoting}
\Name{Meng Y., McAvoy A. \and Li A.} \REVIEW{Proc. Natl. Acad. Sci. U.S.A.}{122}{2025}{e2509575122}.

\bibitem{wang2026strategy}
\Name{Wang X., Zhou L., McAvoy A., Tian Z. \and Li A.} \REVIEW{Proc. Natl. Acad. Sci. U.S.A.}{123}{2026}{e2516380123}.

\bibitem{li2026temporality}
\Name{Li A., Meng Y., Zhou L., Masuda N. \and Wang L.} \REVIEW{Nat. Commun.}{17}{2026}{6238}.

\bibitem{wang2025emergence}
\Name{Wang D., Yi P., Hong Y., Chen J. \and Yan G.} \REVIEW{PLoS Comput. Biol.}{21}{2025}{e1012891}.

\bibitem{meng_y_nc24}
\Name{Meng Y., Cornelius S.~P., Liu Y.-Y. \and Li A.} \REVIEW{Nat. Commun.}{15}{2024}{3125}.

\bibitem{zeng2025evolutionary}
\Name{Zeng Z., Feng M. \and Szolnoki A.} \REVIEW{IEEE Trans. Netw. Sci. Eng.}{13}{2026}{296}.

\bibitem{li2025collective}
\Name{Li Z., Zhang J., Chen X. \and Szolnoki A.} \REVIEW{IEEE Trans. Artif. Intell.}{7}{2026}{1143}.

\bibitem{su2025evolutionary}
\Name{Su Q. \and Stewart A.~J.} \REVIEW{Nat. Commun.}{16}{2025}{4023}.

\bibitem{taylor2007evolution}
\Name{Taylor P.~D., Day T. \and Wild G.} \REVIEW{Nature}{447}{2007}{469}.

\bibitem{tarnita_jtb09}
\Name{Tarnita C.~E., Ohtsuki H., Antal T., Fu F. \and Nowak M.~A.} \REVIEW{J. Theor. Biol.}{259}{2009}{570}.

\bibitem{tarnita2011multiple}
\Name{Tarnita C.~E., Wage N. \and Nowak M.~A.} \REVIEW{Proc. Natl. Acad. Sci. U.S.A.}{108}{2011}{2334}.

\bibitem{mcavoy2022evaluating}
\Name{McAvoy A. \and Wakeley J.} \REVIEW{Proc. Natl. Acad. Sci. U.S.A.}{119}{2022}{e2119656119}.

\bibitem{wu2013dynamic}
\Name{Wu B., Traulsen A. \and Gokhale C.~S.} \REVIEW{Games}{4}{2013}{182}.

\bibitem{mcavoy2016structure}
\Name{McAvoy A. \and Hauert C.} \REVIEW{J. Math. Biol.}{72}{2016}{203}.

\bibitem{su_q_njp16}
\Name{Su Q., Li A., Zhou L. \and Wang L.} \REVIEW{New J. Phys.}{18}{2016}{103007}.

\bibitem{su_q_njp17}
\Name{Su Q., Li A. \and Wang L.} \REVIEW{New J. Phys.}{19}{2017}{103023}.

\bibitem{su2018evolution}
\Name{Su Q., Li A. \and Wang L.} \REVIEW{J. Theor. Biol.}{442}{2018}{149}.

\bibitem{su2019edge}
\Name{Su Q., Zhou L. \and Wang L.} \REVIEW{PLoS Comput. Biol.}{15}{2019}{e1006947}.

\bibitem{svoboda2024density}
\Name{Svoboda J. \and Chatterjee K.} \REVIEW{Proc. Natl. Acad. Sci. U.S.A.}{121}{2024}{e2405605121}.

\bibitem{wang2024deterministic}
\Name{Wang X., Fu F. \and Wang L.} \REVIEW{J. R. Soc. Interface}{21}{2024}{20240055}.

\bibitem{wang2025evolutionary}
\Name{Wang D., Yi P., Yan G. \and Fu F.} \REVIEW{IEEE Trans. Netw. Sci. Eng.}{13}{2025}{5074}.

\bibitem{xia2025behavioral}
\Name{Xia Y., McAvoy A. \and Su Q.} \REVIEW{Natl. Sci. Rev.}{12}{2025}{nwaf403}.

\bibitem{ohtsuki2006evolutionary}
\Name{Ohtsuki H. \and Nowak M.~A.} \REVIEW{Proc. R. Soc. B}{273}{2006}{2249}.

\bibitem{altrock2017evolutionary}
\Name{Altrock P.~M., Traulsen A. \and Nowak M.~A.} \REVIEW{Phys. Rev. E}{95}{2017}{022407}.

\bibitem{wu2023evolutionary}
\Name{Wu T., Fu F. \and Wang L.} \REVIEW{J. Autom. Intell.}{2}{2023}{79}.

\bibitem{shao2024value}
\Name{Shao S. \and Wu B.} \REVIEW{Chaos}{34}{2024}{123128}.

\bibitem{gao2024evolutionary}
\Name{Gao M., Li Z. \and Wu T.} \REVIEW{Chaos}{34}{2024}{113120}.

\bibitem{koutis2014approaching}
\Name{Koutis I., Miller G.~L. \and Peng R.} \REVIEW{SIAM J. Comput.}{43}{2014}{337}.

\bibitem{szolnoki_prx13}
\Name{Szolnoki A. \and Perc M.} \REVIEW{Phys. Rev. X}{3}{2013}{041021}.

\bibitem{roca2009evolutionary}
\Name{Roca C.~P., Cuesta J.~A. \and S{\'a}nchez A.} \REVIEW{Phys. Life Rev.}{6}{2009}{208}.

\bibitem{szolnoki_njp16}
\Name{Szolnoki A. \and Perc M.} \REVIEW{New J. Phys.}{18}{2016}{083021}.

\bibitem{fu2009evolutionary}
\Name{Fu F., Wang L., Nowak M.~A. \and Hauert C.} \REVIEW{Phys. Rev. E}{79}{2009}{046707}.

\bibitem{zhang2011ambivalent}
\Name{Zhang H., Gao M., Li Z., Maa Z. \and Wang H.} \REVIEW{Physica A}{390}{2011}{1961}.

\bibitem{jiang2011effects}
\Name{Jiang L.-L., Zhou T., Perc M. \and Wang B.-H.} \REVIEW{Phys. Rev. E}{84}{2011}{021912}.

\bibitem{wang2023high}
\Name{Wang H., Li A. \and Wang L.} \REVIEW{Chaos Solitons Fractals}{173}{2023}{113602}.

\end{thebibliography}

\end{document}